\documentstyle[12pt]{article}

\def\beq{\begin{equation}}   \def\eeq{\end{equation}}

\newcommand{\gsim}{\lower.7ex\hbox{$\;\stackrel{\textstyle>}{\si
m}\;$}}
\newcommand{\lsim}{\lower.7ex\hbox{$\;\stackrel{\textstyle<}{\si
m}\;$}}

\newcommand{\ra}{\rightarrow}

\begin{document}

\begin{titlepage}
\renewcommand{\thefootnote}{\fnsymbol{footnote}}

\begin{flushright}
TPI-MINN-96/29T\\
UMN-TH-1525/96\\
hep-th/9612490\\

\end{flushright}

\vspace{0.3cm}

\begin{center}
\baselineskip25pt

{\Large\bf 
A More Minimal Messenger Model of Gauge-Mediated 
Supersymmetry 
Breaking?}

\end{center}

\vspace{0.3cm}

\begin{center}
\baselineskip12pt

\def\thefootnote{\fnsymbol{footnote}}

{\large G. Dvali}

\vspace{0.2cm}
Theory Division, CERN, CH-1211 Geneva 23, Switzerland
\vspace{0.2cm}

{\em and}

\vspace{0.3cm}
{\large  M.~Shifman} 

\vspace{0.2cm}
Theoretical Physics Institute, University of Minnesota, Minneapolis, 
MN 55455, USA \\

\vspace{1.5cm}

{\large\bf Abstract} \vspace*{.25cm}
\end{center}
This Letter addresses a provocative question: ``Can the standard 
electroweak Higgs doublets 
and
their color-triplet partners be the messengers of a low energy
gauge-mediated SUSY breaking?" Such a possibility does not seem to 
be immediately  
ruled
out.  If so, it can lead to a very economical scheme with clear-cut 
predictions quite distinct
 from those of the conventional  gauge-mediated scenario. Namely, 
we get
(i) a single light Higgs below the original SUSY-breaking scale; (ii)
tan$\beta = 1$;  (iii) flavor non-universal, but automatically 
flavor-conserving
soft scalar masses; (iv) a light colored scalar with  peculiar
phenomenology. The familiar $\mu$ problem looses its meaning in 
this
approach.

\vspace{2cm}

\hfill

\begin{flushleft}

December 1996

\end{flushleft}

\end{titlepage}

\section{Introduction}

The origin of supersymmetry (SUSY) breaking remains a key 
problem for
 modern models of fundamental interaction. For  phenomenological
implications the most important question, however, is not the precise
nature of supersymmetry breaking {\it per se} but, rather, how this
breaking is communicated to the low-energy observable sector:
quarks, leptons and  gauge fields. Recently  interest has been revived 
\cite{RGM} in  the
so-called ``gauge-mediated" low-energy SUSY
breaking scenarios \cite{gauge}. This approach is motivated 
predominantly by 
its predictivity 
and
a potential  for solving the  problem of the flavor-changing neutral
currents \cite{SDDS}. 
The main ingredients of the approach are as
follows. Supersymmetry breaking occurs through one of the known
 mechanisms (usually, through a dynamical mechanism 
\cite{dynam}) in some 
(usually strongly
coupled) hidden sector of the theory. The  role of this sector
is to provide a $G_W = SU(3)\otimes SU(2)\otimes U(1)$-singlet
chiral superfield $X$ with  non-vanishing vacuum expectation values 
(VEV) of
its auxiliary and scalar components \footnote{Below we will denote 
the
chiral superfields as well as their lower components by the same
symbols. In each case it will be clear from the context to which 
component
we refer to.}
\begin{equation}
\langle X \rangle \neq 0,~~~\langle F_X \rangle \neq 0.
\label{vev}
\end{equation}
The above VEVs ensure
breaking of both supersymmetry
and the $R$ symmetry. A key role in transmitting this breaking from 
the
hidden sector to the visible matter belongs to  the so called 
``messenger" 
sector composed of
the superfields $\phi,~\bar{\phi}$ in the vector-like representations 
of the
standard model (SM)  gauge group. Sometimes this mechanism is 
considered 
in the context of grand unified theories (GUT). In this case, 
$\phi,~\bar{\phi}$ are assumed to belong to the vector-like 
representations of 
the
GUT  gauge group.
The fields in $\phi,~\bar{\phi}$ experience a 
tree level Fermi-Bose mass splitting due to a direct coupling
with the $X$ superfield in the superpotential
\beq
W = hX\bar{\phi}\phi\, . 
\label{xcoupling}
\eeq
A standard ``minimal" choice for the messengers in the GUT version 
is 
$\phi\sim 5$ and 
$\bar{\phi}\sim \bar 5$. Even if grand unification is not considered, 
the 
minimal choice remains essentially the same,  
the components of $\phi,~\bar{\phi}$  are assumed to transform 
under $G_W$ 
as
\beq
\{\bar{3},1,(-2/3)\} + \{1,2,1\}\, ,
 \label{rep}
\eeq
and their conjugates, respectively. This assignment ensures  
non-vanishing
{\it one-loop} soft masses for the ``observable" gauginos
\beq
m_{\lambda_i} \sim K_i{\alpha_i \over 4\pi} \left ({F_X \over 
X}\right
)\, ,
\label{fermions}
\eeq
and {\it two-loop} soft masses for the scalars
\beq
   m_{\lambda_i} \sim K_iC_i\left ({\alpha_i \over 4\pi}\right )^2
\left ({F_X \over X}\right
)^2\, ,
\label{scalars}
\eeq
where $\alpha_i$ are the gauge couplings, and $K_i$ and $C_i$ are
group-theoretical factors which only depend on the gauge quantum 
numbers.
Thus, in the minimal case we can parameterize our ignorance of the 
messenger
sector by a single ratio 
$$
\Lambda = F_X / X
$$ 
which,
for  realistic soft masses,  must be $\sim 100$ TeV or so.

Along with obvious attractive properties,   the  low-energy 
gauge-mediated scenario suffers from an aesthetically ugly feature: 
the 
messenger sector is composed
of {\em ad hoc} fields whose sole {\em raison d'etre} is to connect,
in a SUSY breaking way, the hidden and observable sectors.
These superfields are
not otherwise motivated, and this brings in a certain
degree of arbitrariness in the theory. Moreover,
 they may lead to a serious cosmological difficulty
\cite{dgp}, as the lightest of the messengers tends to be a stable 
particle.
To avoid the problem, an instability must be ensured by
postulating  messenger couplings  to  ordinary matter. Such couplings
may introduce back the flavor non-universality and
unacceptably strong baryon number violation
(unless they  are
strongly suppressed by hand). 

Another serious difficulty of this approach is the $\mu$ problem 
\cite{mu}. One
needs to generate both a supersymmetric Higgs mass term
(Higgsino mass) in the superpotential
\beq
\Delta W = \mu \bar{H}H \label{mu}
\eeq
and a 
soft scalar bilinear mass ($B\mu$  term) in the potential
\beq
\Delta V_{\rm soft} = B\mu\bar{H}H  + \mbox{h. c.}
 \label{Bmu}
\eeq
of the right order of magnitude,
\beq
   B\mu \sim \mu^2 \sim (100 GeV)^2.
\eeq
Usually this is difficult to achieve in the minimal schemes, since, once
forbidden at the tree level, one tends to end up with a problematic 
relation
$B\mu \gg \mu^2$ (for a more detailed discussion and possible ways 
out 
see \cite{mu} and references therein).

In view of the above, a natural question arises: can one exploit the
fields which exist in the modern theory anyway, to assign to them
the messenger role? 
In the present Letter  we address this issue, and analyze whether the
electroweak Higgs doublets $H,\bar H $ and their color-triplet GUT
partners $T, \bar T$ can play
the role of the messengers of the low-energy gauge-mediated SUSY
breaking.

Thus, our task is the search for a
minimal messenger 
model (MMM). 

If $T, \bar T$ can play
the role of the messengers, the aesthetically unpleasant feature
of the approach is eliminated. On the practical side, our minimal 
messenger 
model exhibits no flavor problem. Its
solution is automatic -- the Yukawa coupling
constants are diagonalized simultaneously with the fermion masses.
The
$\mu$ problem gets a different (essentially $no$) meaning, since, as 
shown
below, in the low-energy sector there is a single light Higgs scalar
\beq
h = {H + \bar H^+ \over \sqrt 2}\, .
\label{higgs}
\eeq
(The mass of this particle will be denoted below as $M_h$.)
The orthogonal superposition and  Higgsinos  are heavy. 
Their
masses are of order $\Lambda$, and they decouple. This does not 
lead to the
usual naturalness problems, however, since the Higgs mass is only 
two-loop
corrected.

Below we will argue that such a scenario is not ruled out and
 leads to  clear-cut predictions, which can be tested in present
and future experiments. Apart from  a very different low-energy 
Higgs
spectrum mentioned above, MMM implies  the existence of  light
color-triplet scalar Higgs particles $T$, whose mass can be close to 
$M_h$.
Usually the light color-triplet   Higgs particles are not considered 
because 
of the
menace of a fast proton decay. In supersymmetric theories
the proton decay can be naturally suppressed, however,
by the so-called Clebsch-factor mechanism, see Ref. \cite{lt}.
Depending on the Clebsch factors emerging in  the underlying
GUT, there are three possible outcomes: the light Higgs triplet decays 
(i)
 only in the quark channels; (ii) only in the
lepton channels; (iii)  appears to be  stable in  the detector and must 
be
observed 
in the form
of  stable charged or neutral hadrons (more exactly, it is not 
absolutely stable,
but the lifetime is large).

\section{Higgs weak doublet and colour-triplet as messengers}

Since we want the ordinary Higgs doublets and their color-triplet 
partners
to be the only messengers of the low-energy gauge-mediated SUSY 
breaking,
we assume  the source of their masses to be a
 coupling with the $X$ superfield in the superpotential, with a
VEV of order $\Lambda$. At first sight,  this sounds impossible,
since  such a light color triplet is believed to
lead to unacceptably fast proton decay. However, generically this is
not true \cite{lt}. The proton decay can be eliminated by 
Clebsch factors; these are  certain dimensionless combinations
of the GUT Higgs VEVs which control the strength of the effective
Yukawa coupling constants after the GUT symmetry breaks down.
These effective Clebsch factors are low-energy remnants of the
heavy sector integrated out at the GUT scale. They can naturally 
decouple
the
Higgs triplet $T$ from some (or all) of the species of quarks and 
leptons,
thus automatically suppressing the proton decay.

 Let us briefly discuss the main idea of the Clebsch-suppression 
mechanism.
Consider a grand unified group $G$ with quarks and leptons 
transforming
in the irreducible (or reducible) representation $\Psi^\alpha$ with
$\alpha = 1,2,3$
being a family index. Let $\Sigma$ denote the Higgs 
representation(s) 
that
beak(s) $G$ to $G_W$ (generically, there can be more than  one
such  Higgs field). Let Higgs doublet and triplet be placed in the
irreducible representation $R$. Then the masses of the ordinary 
fermions
are generically induced from a set of  effective $G$ invariant 
operators
(a $G$ invariant contraction of the group indices is assumed)
\beq
R\left ({\Sigma \over M_G}\right )^{n_{\alpha,\beta}}
\Psi^{\alpha}\Psi^{\beta}\, .
 \label{bma}
\eeq
These operators  are induced after  integrating out
all  heavy fields at $M_G$
and, depending on the  precise structure of the theory up there,
may have different  contractions of the 
indices and different   flavor dependence. 
In  view of the fact that none of the minimal SUSY GUTs (e.g. 
$SU(5)$ or
$SO(10)$) with the minimal Yukawa interactions (corresponding to 
$n = 
0$
in Eq. (\ref{bma}))  can account for the observed pattern of the 
fermion 
masses,
such operators are very much motivated.
An important consequence of such a construction is that, after 
breaking the
GUT symmetry by  the vacuum expectation value of $\Sigma$, the 
universality of the resulting
doublet and triplet Yukawa coupling constants is generically 100$\%$ 
violated.
The relation between the couplings is
determined by the group theoretical (Clebsch) factors.
It is perfectly natural that for a certain  choice of the above 
operators
the triplet  turns out to be decoupled from some (or all) species of 
the 
quark and lepton superfields. In  other words, the  triplet  is coupled 
to
matter only in combination with certain   components of $\Sigma$ 
and is
automatically decoupled if the  latter have vanishing  (or small) 
expectation 
values. In short, the Clebsch mechanism \cite{lt} insures the 
decoupling of the
triplet   Higgs not by adjusting its mass to be huge, as in the 
standard scenario
\cite{Moh}, but, rather, through suppressing the corresponding 
coupling constants.
It  exhibits certain  advantages over the
standard doublet-triplet  mass-splitting solutions \cite{Moh}, since it 
kills
simultaneously both dimension-5 \cite{d5}
and dimension-6 proton decay-mediating
operators.
For  further  details
the reader  is referred to  Ref. \cite{lt}. Here we simply parameterize  
the 
Clebsch
factors by independent parameters subject to the constraint of the
 proton stability, and then consider phenomenologically the most 
promising 
possibilities.

Consider first the Higgs spectrum. By assumption, the states
$H,\bar {H}$  and $T,\bar {T}$  get   supersymmetric and
non-supersymmetric contributions to their masses from the 
couplings
to the $X$ field in the superpotential
\beq
 W = gX\{ \bar{H}H + (1 + a) T\bar T\} \, .
\eeq
To allow for different relative wave function
renormalizations  from the GUT scale down to the scale of the 
messenger
sector we have introduced above a factor
 $a$. Its value will be discussed shortly. 
After supersymmetry breaking takes place the above superpotential 
leads to 
the
following tree-level masses
\beq
 g^2|X|^2 \left ( |H|^2 + |\bar H|^2 + (1 + a)^2(|T|^2 + |\bar T|^2) \right )
+ \left [ gF_X \left (H\bar H + (1 + a)T\bar T \right ) +  \mbox{h.c.} 
\right ]\, .
\eeq
If we assume, for definiteness, that $F_X < 0$,  the condition of  
existence of a 
light doublet is
\beq
g^2|X|^2 + gF_X = M^2_h \ll \Lambda^2\, .
 \label{condition}
\eeq
Then the combination indicated in Eq.  (\ref{higgs}) is the lightest 
Higgs,
with 
 the mass $M_h$, while the orthogonal combination $\sim
(H-\bar H^+)$ is heavy, with the mass 
 $\sim g|X|$.
The lightest scalar triplet is given by a similar combination,
\beq
T_h = {T + \bar T^+ \over \sqrt 2}\, .
\label{thiggs}
\eeq
In fact, the most obvious choice is $M^2_h = 0$ (at the scale 
$\Lambda$).
Some ideas as to how this cancellation may naturally take place  will 
be 
discussed shortly. Here we want to mention that, even being  
regarded as an 
explicit
input fine-tuning, Eq. (\ref{condition}) still is a much less severe 
condition than the fine-tuning in the standard
 $SU(5)$ \cite{Moh}, since here we fine-tune two orders of 
magnitude {\em 
versus} 14 orders in the standard version.

Another advantage over the standard  $SU(5)$ approach, where the 
fine-tuning (to zero)  does
exhibit the $\mu$ problem in the gauge-mediated supersymmetry 
breaking
framework, is that our suggestion eliminates the  $\mu$ problem 
altogether.

\subsection{Minimal $SU(5)$}

This is the most economic version. We have only $T$'s and $H$'s
as messengers of SUSY breaking. The mass of Eq. (\ref{thiggs}), 
however,
generically is substantially heavier than $M_h$, see
Eq. (\ref{tmass}) below. This is
due to the fact that the breaking $SU(5)\ra G_W$ occurs at a very 
high scale,
and evolving down to $\Lambda$ brings in the difference
in the wave function renormalizations. Generically, the constant $a$ 
is several 
units. Although this does not preclude the messengers from their
mission of SUSY breaking,  other potentially appealing features
appearing due to $M_h\sim M_{T_h}$, discussed in Sect.  3, are lost.
Note, however, that $T$ can have a large Yukawa coupling with some 
of 
the 
species of
the third generation, of order unity.
The wave function renormalization   due to the
gauge coupling and due to the Yukawa couplings have opposite signs,
and tend to cancel each other. It my happen that, thanks to this 
cancellation, 
$a$ is numerically  rather  small, and $M_h\sim M_{T_h}$ is still
valid.

\subsection{Advanced GUT's}

Now let us briefly discuss how the  cancellation (\ref{condition})  
may happen 
due to
 symmetries of the theory. The same symmetries will ensure also 
that
$a\ll 1$.
In this paper we would not like to enter into  a detailed discussion of 
specific models;  an  example we give must be rather regarded  as an
``existence proof".  

The cancellation  may be ensured  by a
pseudo-Goldstone nature \cite{Gold} of $T$ and $H$ \footnote{The 
idea was
used previously, in particular,  for the
interpretation of the solution in Ref. 
\cite{mu}.
Moreover, it was pointed out,  in a different context \cite{aph},  that 
it can 
ensure 
the
cancellation even for large $\mu$ and $B\mu$ parameters.}.
Imagine that the GUT symmetry is $SU(6)$; it is 
broken to $G_{\rm intrmdt}
 = SU(3)\otimes SU(3)\otimes U(1)$ at some high 
scale $M_G$ and then, at
a lower  scale $M$, not far from $\Lambda$, is further broken to 
$G_W$.
The sector of the theory responsible for the breaking
of $SU(6) $ down to $G_W$  contains a $35$-plet and $6$-  
($\bar 6$)-plets.
Assume that $H$ ($\bar H$) and $T$ ($\bar T$)
states belong to the fundamental $6$-plet ($\bar 6$-plet) of
$SU(6)$. To this end they should be supplemented by an additional
$G_W$-singlet field $S$ ($\bar S$). The embedding is such that
$H$ and $S$ compose a triplet under one of the $SU(3)$ subgroups,
while $T$ is a triplet under another $SU(3)$ subgroup. 

With the gauge interaction switched off, and with both $6$-  and 
$\bar 6$-plets
developing VEV's, we would get twice more ``phase" fields than the 
number
of such fields that are
actually eaten in the (super)Higgs mechanism. Combinations 
orthogonal to 
those
that are eaten in the Higgs mechanism remain massless at the tree 
level.
Their masses appear only after SUSY breaking and are small. If we 
additionally
assume that the VEV of the $(T,H,S)$ $6$-plet are significantly 
smaller
than that of the ``other" $6$-plet, then the pseudo-Goldstone bosons
discussed above will be almost pure $(T,H,S)$.

A relevant superpotential
can be written as
\beq
 W = gX\left ((1 + a)T\bar T + H \bar H +  S\bar S - M^2_1 \right ) \, .
\eeq
In the  example at hand $a = 0$ (above the scale $M$) if no
additional states are introduced at $M_G$.
The factors of the relative wave function renormalization of 
doublet(s) and
singlet(s) are equal because of the $SU(3)$ symmetry. The equality
of the relative wave function renormalization of $T$ and $H,S$
(i.e. $a = 0$) is due
to an obvious  symmetry of the theory under the interchange of two 
$SU(3)$ groups. A very small value of $a$ is only generated
below the scale $M$ of the second breaking
\footnote{This scheme above can exhibit a  difficulty  in 
accommodating the 
standard unification of the gauge couplings. This problem may be 
solved
by a suitable extension of the field content.
We thank I. Gogoladze for bringing such examples to our attention.}.

Now,  SUSY  is spontaneously broken  whenever
dynamics induces a non-vanishing  VEV of $X$. Indeed the
minimization with respect to all
other fields {\it automatically} gives Eq. 
(\ref{condition}) provided $X \neq 0$. This fact is not surprising. 
Indeed, the 
state 
(\ref{higgs})
appears to be the  pseudo-Goldstone of the broken  $SU(3)$ 
and, thus, must be massless at the tree level. At  the same time,  the
mass of the lightest scalar triplet  state (\ref{thiggs}) is given
by
\beq
m_T^2 = g^2 |X|^2 a(1+a) \, . 
\label{tmass}
\eeq
It vanishes in the limit $a\ra 0$, as it should.
The precise value of $a$ depends on  details of the GUT  scheme in 
question.
The most interesting phenomenology results for $a \ll 1$ which we 
briefly
discuss below. 

\subsection{Soft mass terms of matter}

Now, let us consider how the soft masses of the matter
fields are generated through our gauge-mediated scenario.
As usual, there are one-loop gaugino (\ref{fermions}) and flavor-
universal
two-loop scalar (\ref{scalars}) soft masses.
A new  crucial point is the generation of the soft scalar masses
at {\em one-loop} level. These mass terms 
are generated due to  direct Yukawa couplings of our Higgs 
messengers
to the squarks and sleptons and are {\em not} flavor-universal. They 
have the 
form
\beq
       m^2_{\alpha\beta} = {G_{\alpha\gamma}^*
G_{\gamma\beta} \over 16\pi^2} O( \Lambda^2)\, ,
 \label{nonu}
\eeq
where $G_{\alpha\beta}$ are the Yukawa coupling constants, and we 
have 
taken into 
account
the fact of SUSY breaking in the Higgs sector, with the scale $O( 
\Lambda )$. 
These masses are manifestly
non-universal; by far the largest appears in the third generation.
What is remarkable, the large and flavor-non-universal squark
masses
 do not lead to flavor violation since they
are diagonalized simultaneously with the fermion masses.
More explicitly the one-loop soft scalar masses for the different 
species
$Q^{\alpha}, d_c^{\alpha}, u_c^{\beta},
L^{\alpha},e_c^{\beta}$ are proportional to:
$$
 m_{Q\alpha\beta}^2 \propto G_{\alpha\gamma}^{*u} 
G_{\gamma\beta}^u + 
G_{\alpha\gamma}^{*d} G_{\gamma\beta}^d,~~~
 m_{u\alpha\beta}^2 \propto G_{\alpha\gamma}^{*u} 
G_{\gamma\beta}^u\, ,
$$
\beq
 m_{d\alpha\beta}^2 \propto G_{\alpha\gamma}^{*d} 
G_{\gamma\beta}^d,~~~
 m_{e\alpha\beta}^2 \propto G_{\alpha\gamma}^{*e} 
G_{\gamma\beta}^e,~~~
 m_{L\alpha\beta}^2 \propto G_{\alpha\gamma}^{*e} 
G_{\gamma\beta}^e,
\eeq
where $G^{Q,u,d,L,e}$ are the respective Yukawa coupling constants.
Unsurprisingly, 
the
flavor violation is suppressed in this scheme and is controlled by the
CKM mixing angles.
Note that the parameter  tan$\beta$ is  extremely  close to 1, and the
Higgs sector essentially looks as that of the standard model.
 Since the above soft masses are
one-loop induced, for the third family our new contribution will
dominate over the usual gauge-mediated two-loop mass. Thus, the
above approach leads to: (i) the hierarchical, but flavor conserving
(aligned), pattern of the soft masses;  (ii) a single light Higgs doublet
$h$; (iii) a light color-triplet  Higgs with mass $\sim a\Lambda$.
 Since the corrections to $M_h$ appear only at two-loop level the
two-loop hierarchy $\Lambda \gg M_h$ is natural.

\section{Light color-triplet Higgs: phenomenological implications}

As was mentioned,  the most interesting phenomenological 
consequences 
will take 
place for
$a \ll 1$.  In this case the light color-triplet scalar can be the
subject of experimental study at existing facilities or those  planned 
for 
the near future. To ensure the
proton stability we assume that the couplings of the Higgs triplet to 
the 
matter
is suppressed by the Clebsch factor via the mechanism of Ref. 
\cite{lt}.
For simplicity we  parametrize the couplings of triplets
with the matter superfields in terms of the flavor independent
Clebsch factors (more complicated versions, with flavor-dependent 
structures,
are also possible). Then the most general Yukawa couplings of the
Higgs triplets in the superpotential are
\beq
 T G_{\alpha\beta}^u\left(C_{ue}u_c^{\alpha}e_c^{\beta}
+ C_{QQ}Q^{\alpha}Q^{\beta}\right ) +
\bar{T} G_{\alpha\beta}^d\left(C_{du}d_c^{\alpha}u_c^{\beta}
+ C_{QL}L^{\alpha}Q^{\beta}\right )\, ,
\eeq
where $C_{ue}, ...$ are the Clebsch factors. To ensure the proton 
stability,
some of these factors (or all of them) must be suppressed. For 
example,
it is enough to have  $C_{QL} = C_{ue} = 0$.
In this case the light triplet, once produced, will tend to decay
into the quark-quark pairs. Note that it is phenomenologically 
impossible to 
allow both the 
quark-quark and quark-lepton decay channels,
since this would lead to unacceptable proton decay.

If so, these Higgs color triplets must be carefully considered
as possible candidates for the ALEPH four-jet  events.
As well known, these events continue to accumulate, on the one 
hand,
and continue to defy any reasonable explanation, on the other.
Phenomenologically, the light triplet  Higgses in this aspect
will look similar to
the down right-handed quarks  in the models with
the $R$ parity violation (for a review of such models see
\cite{bhat,cglw}).
Phenomenology of the two-quark decays is the same.
If the $R$ parity violation explanation goes through (see e.g. 
\cite{RPV}),
the same should be valid for the light triplet  Higgses. The opposite is 
also true
\footnote{This remark may not apply to the explanation
of Ref. \cite{cglw}, as this work  does not assume the pair production
but, rather, assumes production of left-handed plus right-handed  
selectrons.}.
Theoretically, there are two important distinctions, however.
First, since we do not violate the $R$ parity, the stability of LSP
is preserved. Second, although (\ref{thiggs}) mimics the
right-handed down squarks, their fermion partners are much 
heavier, with 
masses of order $\Lambda$, in sharp contradistinction  to  the
situation with the
right-handed down squarks, whose fermion partners are lighter than
the squarks themselves. 

Two further theoretical points deserve mentioning. If the triplet 
Higgs is
light, it gives, through a loop, a noticeable contribution to $Z\ra b\bar 
b$
yield, roughly at the level of that associated with the light stop and 
chargino
(see e.g. Ref. \cite{Stefan}). This puts the theoretical prediction 
for $R_b$ right on top of the existing world average for $R_b$, which 
is 
slightly higher than the standard model prediction. Simultaneously,
the genuine value of 
$\alpha_s (M_Z)$ goes down to 0.112, which is also welcomed 
\cite{EKSV}.

On the other hand, the light triplet  Higgs spoils unification of
the gauge couplings within $SU(5)$ GUT. Given the experimental 
values
of $\alpha_1$ and $\alpha_2$ we get a value of $\alpha_3$ too 
low
to be compatible with data.

What if all Clebsch factors vanish in the supersymmetric limit? Their
typical value after supersymmetry breaking, induced due to the shift
of the heavy VEVs, is ${\alpha \Lambda  / 4\pi M_G}$ \cite{lt}.
This is certainly not enough to mediate (an observable) proton decay
or to make triplets decay in the detector. Therefore, experimentally  
such a
decoupled scalar triplet  should be observed in the form  of  stable
possibly charged hadrons. (They are not truly stable, 
and can not accumulate in matter, but rather the lifetime is large.)
Phenomenology of such states will be somewhat similar to that 
discussed in \cite{so10} in a different  context.

\subsection{Conclusions and Outlook}

In this Letter  we have suggested the possibility that the 
standard
electroweak Higgs doublets and their color-triplet partners are  the
messengers of a gauge-mediated low energy supersymmetry 
breaking.
While {\em a priori} it is not obvious that such a possibility is 
free of inconsistencies,
it does not seem to be ruled out so far. If so,   a very
economical, predictive and exciting scenario may emerge. The MMM 
approach, in its simplest form,
is quite restrictive and leads to clear-cut predictions
different from those of more conventional messenger 
scenarios:

1) a single light Higgs scalar
below the scale $\Lambda$, in particular, implying ${\rm tan}\beta = 
1$;

2) flavor non-universal, but automatically flavor-conserving
soft scalar masses; 

3) the  possibility of the light scalar Higgs triplet (with
 quantum numbers of $d_c$) decaying only into quarks (only into
leptons), or not decaying at all (in detector).
The  phenomenology of such  colored scalar may be a 
subject
of speculation in connection with the recent ALEPH four-jet  events.

Potential difficulties of the above approach, which may require
further assumptions about physics above the scale $\Lambda$,
are  related to  the gauge coupling unification. Also, the
electroweak symmetry breaking in this scheme deserves a careful 
study.

\vspace{0.3cm}

{\bf Acknowledgments}: \hspace{0.2cm} 

We are grateful to Gian Giudice for very useful comments and 
discussions. One of the authors (M.S.) would like to thank
the CERN Theory Division, where this work began, for its kind 
hospitality.

This work was supported in part by DOE under the grant number
DE-FG02-94ER40823.



\begin{thebibliography}{99}

\bibitem{RGM}
Some recent works are: M. Dine and A. Nelson, {\it Phys. Rev. } {\bf 
D47}
(1993) 1277; M. Dine, A. Nelson, and Y. Shirman, {\it Phys. Rev. } {\bf 
D51}
(1995) 1362;  M. Dine, A. Nelson, Y. Nir, and Y. Shirman, {\it Phys. 
Rev. }
{\bf 
D53} (1996) 12658.
For a brief review see A. Nelson, hep-ph/9511218.
A detailed discussions of phenomenological implications
is given e.g. in  S. Dimopoulos, S. Thomas, and J. D. Wells, 
hep-ph/9609434.

\bibitem{gauge} 
M. Dine, W. Fischler and M. Srednicki, {\it Nucl. Phys. } {\bf B189}
(1981) 575;
S. Dimopoulos and S. Raby, {\it Nucl. Phys.} {\bf B192}
(1981) 353; 
M. Dine, W. Fischler, {\it Phys. Lett. } {\bf B110} (1982) 227;
M. Dine and M. Srednicki, {\it Nucl. Phys. } {\bf B202}
(1982) 238;
M. Dine and  W. Fischler, {\it Nucl. Phys. } {\bf B204} (1982) 346;
L. Alvarez-Gaum\'e, M. Claudson and M. Wise,
{\it Nucl. Phys. } {\bf B207}
(1982) 96;
C.R. Nappi and B.A. Ovrut, {\it Phys. Lett. } {\bf B 113} (1982) 175;
S. Dimopoulos and S. Raby, {\it Nucl. Phys.} {\bf B219} (1983) 479.

\bibitem{SDDS}
The essence of the problem is presented in a concise but exhaustive 
form for example in 
S. Dimopoulos and D. Sutter, {\it Nucl. Phys.} {\bf B452} (1995) 496.

\bibitem{dynam}
The dynamical mechanism was pioneered by
I. Affleck, M. Dine, and N. Seiberg, {\it Nucl. Phys.} {\bf B256} (1985)
557; for a review see A.I. Vainshtein, V.I. Zakharov and M.A. 
Shifman,
{\it Usp. Fiz. Nauk} {\bf 146} (1985) 683
[{\it Sov. Phys. - Uspekhi} {\bf 28} (1985) 709].

\bibitem{dgp}
S. Dimopoulos, G. Giudice and A. Pomarol, hep-ph/9607225.

\bibitem{mu} 
G. Dvali, G. Giudice and A. Pomarol, {\it Nucl. Phys.} {\bf B478} (1996)
31.
 
\bibitem{lt}
G. Dvali, {\it Phys. Lett.} {\bf B287} (1992) 101;
{\it Phys. Lett.} {\bf B372} (1996) 113;


\bibitem{Moh}
See e.g. the textbook
R.N. Mohapatra, {\it Unification and Supersymmetry},  2-nd Edition, 
(Springer-Verlag, Berlin, 1992), Chapter 13. 

\bibitem{d5} 
N. Sakai and T. Yanagida, {\it Nucl. Phys.} {\bf B197} (1982) 533; \\
S. Weinberg, {\it Phys. Rev.} {\bf D26} (1982) 287; \\
M. Vysotsky, {\it Phys. Lett.} {\bf B114} (1982) 125;
T.M. Aliev and M. Vysotsky, {\it Phys. Lett.} {\bf B120} (1983) 119.

\bibitem{Gold} See e.g.,
K. Inoue, A. Kakuto and T. Takano, {\it Prog. Theor. Phys.} {\bf 75}
(1986) 664;
A. Anselm and A. Johansen, {\it Phys. Lett.} {\bf B200} (1988) 331;
A. Anselm, {\it Sov. Phys. JETP} {\bf 67} (1988) 663;
Z. Berezhiani and G. Dvali, {\it Sov. Phys.} Lebedev Institute Reports 
{\bf 5}
(1989) 55; R. Barbieri, G. Dvali and M. Moretti, {\it Phys. Lett.} {\bf 
B312}
(1993) 137.

\bibitem{aph} 
A. Pomarol, private communications and hep-ph/9608230 (Talk 
given at SUSY-96);
S. Pokorski and R. Hempfling, private communications.

\bibitem{bhat}
G. Bhattacharyya, hep-ph/9608415.

\bibitem{cglw}
Recently an interesting explanation has been suggested, that does not
imply a pair production,
M. Carena, G.F. Giudice, S. Lola and C.E.M. Wagner,
hep-ph/9612334.

\bibitem{RPV} 
A. Grant, R. Peccei, T. Veletto and K. Wang,
{\it Phys. Lett.} {\bf  B379} (1996) 272; D. Choudhury and D. Roy,
{\it Phys. Rev. } {\bf D54} (1996) 6797; 
H. Dreiner, S. Lola, and  P. Morawitz, hep-ph/9606364;
K. Ghosh, R. Godbole, and S. Raychaudhury, hep-ph/9605460;
P. Chankowski, D. Choudhury and S. Pokorski, hep-ph/9606415.

\bibitem{Stefan} 
P. Chankowski  and S. Pokorski, {\it Nucl. Phys. } {\bf B475} (1996) 3;
J. Wells and G. Kane, {\it Phys. Rev. Lett.} {\bf 76} (1996) 4458;
W. de Boer, A. Dabelstein, W. Hollik, and W. Mosle, hep-ph/9607286;
J. Ellis, J. Lopez and D. Nanopoulos, hep-ph/9612376.

\bibitem{EKSV}
S. Eidelman, L. Kurdadze, M. Shifman, and A. Vainshtein,
to be published.

\bibitem{so10} 
R. Barbieri, G. Dvali and A. Strumia,
{\it Nucl. Phys. } {\bf B391} (1993) 487.

\end{thebibliography}
\end{document}